\documentclass[a4paper,fleqn,usenatbib]{mnras}

\usepackage{newtxtext,newtxmath}

\usepackage[T1]{fontenc}
\usepackage{ae,aecompl}

\usepackage{graphicx}	
\usepackage{amsmath}	
\usepackage{amssymb}	
\usepackage{soul}		
\usepackage{multirow}	
\usepackage{pdflscape}	
\usepackage[dvipsnames]{xcolor}

\title[The NGC~7232 galaxy group]{WALLABY Early Science - II. The NGC~7232 galaxy group}

\author[K. Lee-Waddell et al.]{
K.~Lee-Waddell$^{1}$\thanks{E-mail: karen.lee-waddell@csiro.au (KLW)}, B.S.~Koribalski$^{1}$, T.~Westmeier$^{2}$, A.~Elagali$^{2,3,1}$, B.-Q.~For$^{2,3}$,  
\newauthor D.~Kleiner$^{1,4}$, J.P.~Madrid$^{1}$, A.~Popping$^{2,5}$, T.N.~Reynolds$^{2,3,1}$, J.~Rhee$^{2,3,5}$, P.~Serra$^{4}$, 
\newauthor L.~Shao$^{6,7}$, L.~Staveley-Smith$^{2,3}$, J.~Wang$^{6}$, M.T.~Whiting$^{1}$, O.I.~Wong$^{2,3}$, 
\newauthor J.R. Allison$^{8,3}$, S.~Bhandari$^{1}$, J.D.~Collier$^{9,1,10}$, G.~Heald$^{11,3}$, J.~Marvil$^{1}$, S.M. Ord$^{1}$
\\
$^{1}$CSIRO Astronomy and Space Science, Australia Telescope National Facility, PO Box 76, Epping NSW 1710, Australia\\
$^{2}$International Centre for Radio Astronomy Research, The University of Western Australia, 35 Stirling Hwy, Crawley WA 6009, Australia\\\
$^{3}$ARC Centre of Excellence for All Sky Astrophysics in 3 Dimensions (ASTRO 3D)\\
$^{4}$INAF - Osservatorio Astronomico di Cagliari, Via della Scienza 5, I-09047 Selargius (CA), Italy\\
$^{5}$ARC Centre of Excellence for All-sky Astrophysics (CAASTRO) \\
$^{6}$Kavli Institute for Astronomy and Astrophysics, Peking University, Beijing 100871, China\\
$^{7}$Research School of Astronomy and Astrophysics, Australian National University, Canberra ACT 2611, Australia\\
$^{8}$Sub-Dept. of Astrophysics, Department of Physics, University of Oxford, Denys Wilkinson Building, Keble Rd., Oxford, OX1 3RH, UK\\
$^{9}$School of Computing, Engineering and Mathematics, Western Sydney University, Locked Bay 1797, Penrith NSW 2751, Australia\\
$^{10}$Inter-University Institute for Data Intensive Astronomy (IDIA), University of Cape Town, Private Bag X3, Rondebosch, Cape Town 7701, South Africa\\
$^{11}$CSIRO Astronomy and Space Sciences, PO Box 1130, Bentley WA 6102, Australia\\
}
\date{Accepted 2018 December 22. Received 2018 December 20; in original form 2018 August 02}

\pubyear{2019}

\begin{document}
\label{firstpage}
\pagerange{\pageref{firstpage}--\pageref{lastpage}}
\maketitle

\begin{abstract}

We report on neutral hydrogen ({H\sc{i}}) observations of the NGC~7232 group with the Australian Square Kilometre Array Pathfinder (ASKAP). These observations were conducted as part of the Wide-field ASKAP L-Band Legacy All-sky Blind surveY (WALLABY) Early Science program with an array of 12 ASKAP antennas equipped with Phased Array Feeds, which were used to form 36 beams to map a field of view of 30 square degrees. Analyzing a subregion of the central beams, we detect 17 {H\sc{i}} sources. Eleven of these detections are identified as galaxies and have stellar counterparts, of which five are newly resolved {H\sc{i}} galaxy sources.  The other six detections appear to be tidal debris in the form of {H\sc{i}} clouds that are associated with the central triplet, NGC~7232/3, comprising the spiral galaxies NGC 7232, NGC 7232B and NGC 7233.  One of these {H\sc{i}} clouds has a mass of M$_{\mathrm{HI}} \sim 3 \times 10^{8}$ M$_{\odot}$ and could be the progenitor of a long-lived tidal dwarf galaxy.  The remaining {H\sc{i}} clouds are likely transient tidal knots that are possibly part of a diffuse tidal bridge between NGC~7232/3 and another group member, the lenticular galaxy IC~5181.

\end{abstract}




\section{Introduction}
\label{sec:intro}

Most galaxies in the current epoch of the Universe reside in galaxy groups \citep{tul1987}. These groups are ubiquitous and their evolution is often dominated by gravitationally driven interactions between their members (\citealt{barweb2001}, \citealt{yan2007}). Neutral atomic hydrogen ({H\sc{i}}) gas can be used to trace disturbed morphologies that are the result of galaxy interactions and detect newly formed tidal features (\citealt{mul2013}, \citealt{lee2016}).  The presence of tidal bridges and tails can constrain the initial properties of the interacting galaxies and the dynamics of the actual encounter (\citealt{too1972}, \citealt{bou2006}), aiding in our understanding of this evolutionary process.  Within the tidal streams, high-density clumps of {H\sc{i}} can accrete sufficient amounts of material to eventually become self-gravitating tidal dwarf galaxies (TDGs; \citealt{mir1992}, \citealt{lel2015}).  TDGs can provide further information about the interaction event and possibly strengthen the standard model of cosmology (see \citealt{duc2014}, \citealt{plo2018}).  

A single galaxy group offers a mere snapshot of this type of evolutionary process.  In order to throughly examine galaxy interactions and fully quantify the by-products of these events, large-scale surveys are required.  Previous single-dish all sky surveys, such as the {H\sc{i}} Parkes All-Sky Survey (HIPASS; \citealt{bar2001}) and Arecibo Legacy Fast ALFA (ALFALFA; \citealt{gio2005}), provided the opportunity to conduct a census of the {H\sc{i}} located in group environments but often higher angular resolution follow-up observations were required to resolve and properly characterize low-mass tidal features (\citealt{ryd2001}, \citealt{eng2010}, \citealt{jan2015}, \citealt{lee2016}).  

The Australian Square Kilometre Array Pathfinder (ASKAP) is a new radio interferometer that will enable wide-field observations that can also resolve low-mass {H\sc{i}} clouds (\citealt{joh2007}, \citealt{joh2008}).  ASKAP consists of 36 $\times$ 12-m antennas equipped with Phased Array Feeds (PAFs) that yield a 30 square degree field of view \citep{deb2009}.  All observations are, by default, conducted using full polarization mode (i.e.~observing in all four Stokes parameters).  The 6-element Boolardy Engineering Test Array (\citealt{hot2014}, \citealt{mcc2016}) has demonstrated the potential of ASKAP as an ideal sky survey instrument (e.g.~\citealt{all2015}, \citealt{ser2015b}, \citealt{hey2016}).  

One of the two top-ranked survey science projects planned for ASKAP is the Wide-field ASKAP L-Band Legacy All-sky Blind surveY (WALLABY; PIs B. Koribalski \& L. Staveley-Smith), which has been designed to study the properties, environments, and large-scale distribution of {H\sc{i}}-rich galaxies.  WALLABY will cover 75 percent of the sky (declination range of $-90^{\circ} < \delta < +30^{\circ}$) and is predicted to detect {H\sc{i}} in more than 500,000 galaxies out to a redshift of 0.26 with an angular resolution of 30 arcsec and a spectral resolution of 4 km s$^{-1}$ (\citealt{duf2012}, \citealt{kor2012}).  This level of resolution is required to resolve the physical characteristics and measure the dynamical properties of tidally formed features in the nearby universe (\citealt{mul2013}, \citealt{lel2015}).

ASKAP Early Science is an observing program aimed at producing scientifically useful data, with at least 12 MkII PAF-equipped ASKAP antennas (i.e.~ASKAP-12), while commissioning ASKAP to its full specification.  The priorities for this program are to demonstrate the unique capabilities of ASKAP, produce data sets to facilitate the development of data processing and analysis techniques, and achieve high scientific impact.  ASKAP survey science teams were given opportunities to select specific science fields to target with ASKAP-12 during the shared-risk and development phases of all aspects of ASKAP.  These observations would not only test the capabilities of the array but also the automated data processing pipeline {\sc{ASKAPsoft}}\footnote{https://data.csiro.au/dap/search?q=ASKAPsoft} as well as the CSIRO ASKAP Science Data Archive\footnote{https://data.csiro.au/dap/public/casda/casdaSearch.zul} (CASDA) for storing the final data products.

The ASKAP Early Science program started in October 2016 with the WALLABY team taking 36-beam observations using ASKAP-12.  The first WALLABY Early Science field, the NGC~7232 field, was chosen because it contains 19 detections from HIPASS (\citealt{kor2004}, \citealt{mey2004}) -- including a nearby spiral galaxy (IC~5201), galaxy pairs, and other interacting systems -- which would assist in data validation.  These varying environments also provide the opportunity to capitalize on ASKAP's high resolution capabilities, allowing for more detailed spatial studies and the potential for new {H\sc{i}} detections.  Although there are several {H\sc{i}}-rich galaxy systems in this field (e.g.~\citealt{rey2018}), this paper will specifically focus on the NGC~7232/3 triplet and its neighbouring galaxies.

The NGC~7232/3 triplet comprises three spiral galaxies, NGC~7232, NGC~7232B and NGC 7233, located in a loose group environment (which is listed as LGG~455 by \citealt{gar1993} and includes a neighbouring lenticular galaxy, IC~5181) at a Hubble distance of $\sim$24 Mpc \citep{gar1995}.  Previous interferometric observations with the Australia Telescope Compact Array (ATCA) show {H\sc{i}} streams connecting the galaxies within the triplet  \citep{barweb2001}, indicating a recent/on-going interaction event.  HIPASS observations show that the triplet system as well as an unresolved starless {H\sc{i}} feature, HIPASS J2214-45, reside in a common gas-rich envelope referred to as the AM2212-460 galaxy group \citep{kor2004}.  There are a handful of other neighbouring galaxies, bringing the membership number to $\sim$10 galaxies \citep{barweb2001}.

Here we are using ASKAP's ability to provide a large field of view while maintaining high angular resolution to carry out detailed analysis of the NGC~7232/3 triplet system and its possible connection to other group members.  Although ASKAP-12 has a limited number of baselines (compared to the full array), longer integration times can be used to increase the {H\sc{i}} sensitivity in order to detect dwarf companions and faint tidal features. Since the angular resolution of ASKAP is comparable to the ATCA observations, the latter provide a good reference for data verification purposes.  The higher frequency resolution of ASKAP will enable further dynamical analysis of spectrally distinct components of the galaxy triplet and its tidal streams.

In Section~\ref{sec:obs} of this paper, we describe the observational setup used during various commissioning and Early Science phases of the ASKAP-12 array.  Section~\ref{sec:processing} details the data processing and imaging completed with {\sc{ASKAPsoft}}.  In Section~\ref{sec:results}, we present our measurements, final {H\sc{i}} maps of the region and stellar properties from ancillary optical imaging.  We analyse and discuss the newly resolved {H\sc{i}} sources in Section~\ref{sec:discuss} and Section~\ref{sec:conclude} contains concluding remarks.

\section{Observations}
\label{sec:obs}

\begin{figure}
\begin{center}
  \includegraphics[width=87mm]{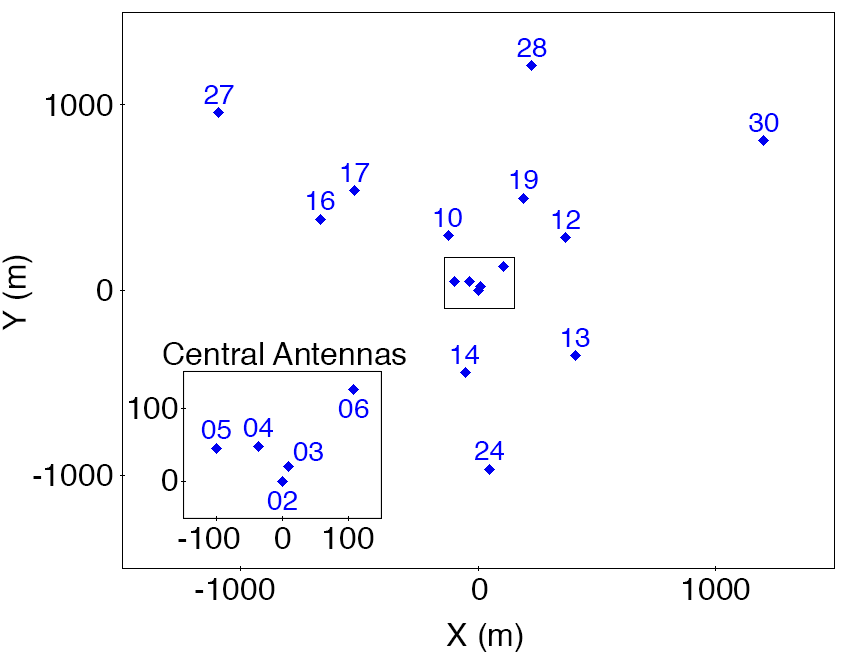}
  \caption{ASKAP positional diagram of antennas used to observe the NGC~7232 field (with antenna 02 set as the centre of the array).  The insert shows the central antennas.  Due to computing limitations during these Early Science and commissioning observations, up to 12 antennas were used for each night of observations (see Tables~\ref{table:array} and \ref{table:obs}).
\label{fig:antennas}}
\end{center}
\end{figure}

The majority of the observations of the NGC~7232 field were taken during the inaugural month of the ASKAP Early Science program.  Additional observations were obtained in August and September during different commissioning phases of the array in 2016 and 2017.  The same configuration of antennas was used throughout 2016 when only 48 MHz of simultaneous bandwidth was achievable.  The bandwidth was split into 2592 independent channels, each 18.5 kHz wide (equivalent to 3.9 km s$^{-1}$ at 1420 MHz).  After numerous system upgrades and further commissioning, more bandwidth and new antennas were available for the main array, which changed the available baselines in 2017 (see Table~\ref{table:array} and Fig.~\ref{fig:antennas}).  Nevertheless, due to computing limitations during Early Science and caution exercised during the later commissioning phases, no more than 12 antennas were used on a given night of observations on the NGC~7232 field. 

\begin{table*}
 \centering
 \begin{minipage}{120mm}
 \caption{ASKAP-12 array configurations used during the observations of the NGC~7232 field}
 \label{table:array}
\begin{tabular}{l c c c c}  
\hline
Date				& Observed		&Antennas available							&Minimum	&Maximum\\
				& frequency range	&										&baseline		&baseline\\
				& (MHz)			&										&(m)			&(m)\\
\hline 
Aug - Oct 2016		& 1376.5 - 1424.5	& 02, 04, 05, 10, 12, 13, 14, 16, 24, 27, 28, 30		&60			&2300\\
Aug 2017			& 1248.5 - 1440.5	& 03, 04, 05, 06, 10, 12, 14, 17, 19, 24, 30		&50			&2120\\
Sep 2017			& 1200.5 - 1440.5	& 02, 03, 04, 06, 10, 14, 16, 17, 19, 27, 28, 30		&20			&2300\\
\hline
\end{tabular} 
\end{minipage}
\end{table*}

\begin{figure}
\begin{center}
  \includegraphics[width=80mm]{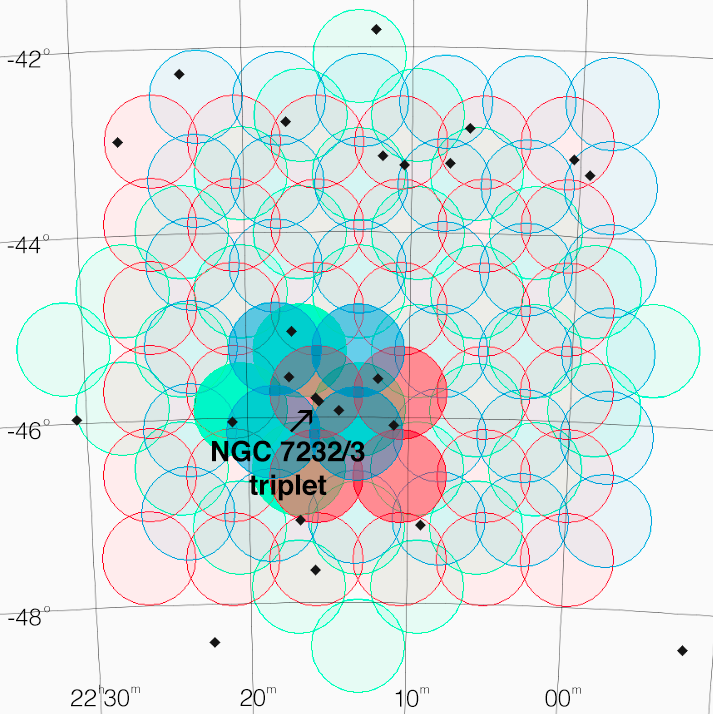}
  \caption{ASKAP 36-beam footprints for the NGC~7232 field observations: red = footprint A, green = rotated footprint A, blue = footprint B (see text for further details).  The black diamonds indicate the location of HIPASS sources.  The darker beams indicate the four beams from each footprint, focused on the NGC~7232/3 triplet, that was processed and imaged for this paper.
\label{fig:footprint}}
\end{center}
\end{figure}

Throughout the Early Science program, science targets were generally observed at night, after ASKAP development and commissioning activities. Table~\ref{table:obs} summarizes all commissioning and Early Science observations on the NGC~7232 field, totalling 18 usable nights.  A square $6 \times 6$ beam footprint, yielding a $5.5 \times 5.5$ degree field of view, was chosen to maximise the sky coverage.  Two interleaves -- footprint A (centred at RA = 22:13:07.7, Dec = -45:16:57.1, J2000) and footprint B (RA = 22:10:35.41, Dec = -44:49:50.7) -- were combined to fill in the gaps between beams and produce a more uniform sensitivity pattern (see Figure~\ref{fig:footprint}).  The data will be publicly available on CASDA under the appropriate ASKAP scheduling block identifying number (SBID; Table~\ref{table:obs}).  \citet{bha2018} have conducted a search for variable and transient continuum sources using footprint B from the same set of observations.  

In September 2016, there was testing of rotating the footprint by 45$^{\circ}$ to better match the configuration of electronics on the PAFs and thereby increase the sensitivity of the corner beams.  The formed beams were not re-rotated resulting in a diamond shape on the sky for those observations.  This rotated beam forming method proved to be beneficial and became standard starting December 2016.  The observations taken in 2017 used re-rotated beams to match the square shape of the October 2016 observations.

Each night began with a few hours of observations on the primary calibrator, PKS1934-638, that was sequentially positioned at the centre of each of the 36 beams for a chosen amount of time per beam ($t_{\mbox{calibrator}}$).  During our Early Science observations, we were able to determine that 200 seconds per beam was sufficient for post-observing calibrations. Longer integration times would significantly increase observing overheads without appreciably improving the calibration procedure. 

\begin{figure}
\begin{center}
  \includegraphics[width=85mm]{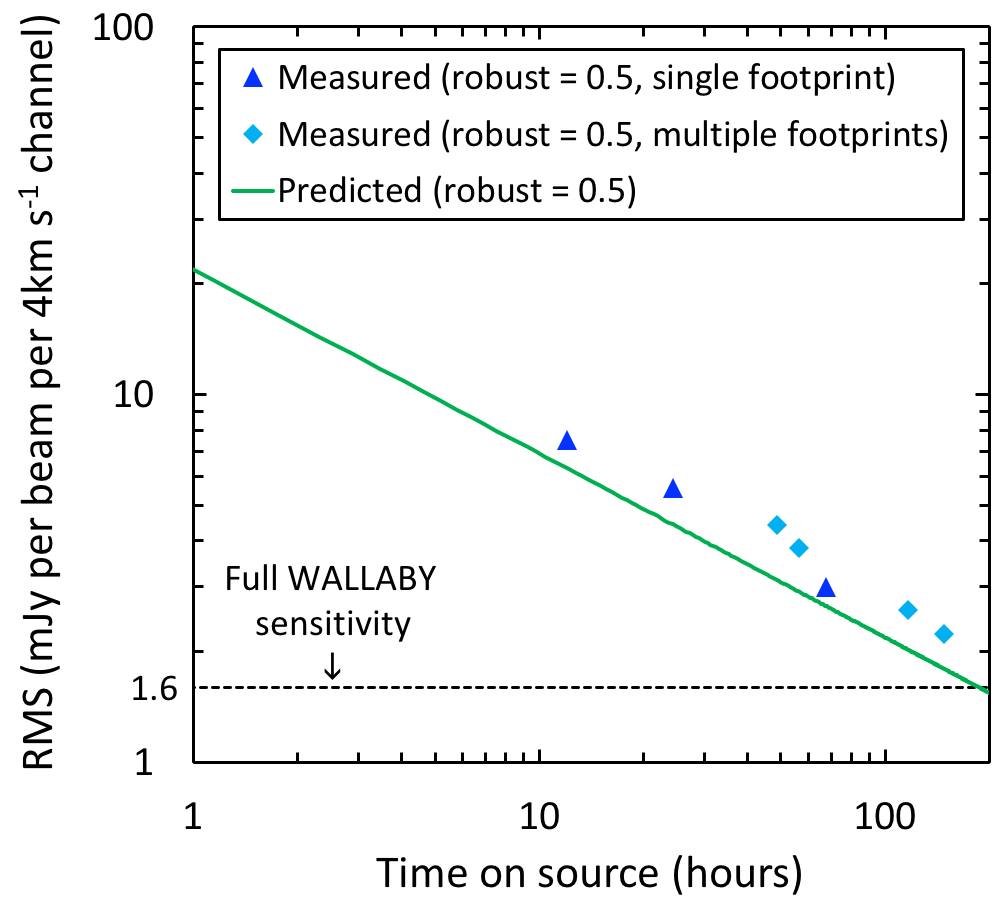}
  \caption{ASKAP per channel RMS noise of the combined image cubes compared to theoretical predictions of ASKAP-12.  Single footprint values are measured from the highest sensitivity region of footprint A and are representative of the channels containing {H\sc{i}} sources.  The noise values for multiple footprints were measured from that same spatial region, which coincides with the highest sensitivity part of those cubes, and includes data from footprint B and the rotated footprint.
\label{fig:rms}}
\end{center}
\end{figure}

Due to technical issues on various nights, which stemmed from commission tests on the overall stability of the array during Early Science, the data from one or two antennas -- as specified in Table \ref{table:obs} -- were omitted from the processing procedure.  The equivalent ASKAP-12 observing time ($t_{\mbox{equivalent}}$) has been computed using the following relation, based on the standard radiometer equation for interferometry:
\begin{equation}
t_{\mbox{equivalent}} = \frac{N(N-1)}{132} t_{\mbox{science}},
\end{equation}
where $N$ is the number of antennas used during each observation, $t_{\mbox{observed}}$ is the actual time spent on the science source and the factor of 132 is from 12*(12-1) total antennas.  The goal was to achieve full WALLABY sensitivity (i.e.~1.6 mJy per beam per channel; \citealt{kor2012}) by combining multiple nights of data.  Out of the total observing time of $t_{\mbox{equivalent}}$ = 180h, 150 hours were used to produce the final image cube that has a sensitivity of $\sim$2.2 mJy per beam per channel (see Section~\ref{sec:processing} for further details). Figure~\ref{fig:rms} shows the ASKAP-12 root-mean-square (RMS) noise measurements and predictions based on a temperature over efficiency value of T$_{\mbox{sys}}$ $\eta^{-1}$ = 85K, which was the value estimated for the telescope at the time of the observations.

\begin{table*}
 \centering
 \begin{minipage}{144mm}
 \caption{ASKAP Early Science observations of the NGC~7232 field. }
 \label{table:obs}
\begin{tabular}{ l c c c c c c c}  
\hline
Obs. date		& ASKAP 		&Footprint		& $t_{\mbox{calibrator}}$	& $t_{\mbox{science}}$	&Omitted 	&$t_{\mbox{equivalent}}$$^\dagger$	& Notes\\
			& SBID		&interleave	& (s)					& (s)					& antennas			& (h)					&\\
\hline 
11 Aug 2016	& 1927		& A			& 200			& 43207.2				& 02, 14			& 8.2					&\\
12 Aug 2016	& 1934		& A			& 200			& 41525.1 			& 14				& 9.6					&\\
5 Sep 2016	& 2012		& A 	 		& 200			& 32348.2				& 02				& 7.5					& footprint rotated 45$^{\circ}$\\
6 Sep 2016	& 2025		& A 			& 200			& 39793.2				& 02				& 11.1				& footprint rotated 45$^{\circ}$\\
7 Oct 2016	& 2238		& B			& 300			& 37354.7				& 14				& 8.6					& excluded from final cube\\
8 Oct 2016	& 2247		& B			& 300			& 41087.1				& 				& 11.4				& excluded from final cube\\\
9 Oct 2016	& 2253		& B			& 300			& 43535.6				& 				& 12.1				& excluded from final cube\\\
10 Oct 2016	& 2264		& B			& 200			& 39604.1				&				& 11.0				&\\
11 Oct 2016	& 2270		& A			& 300			& 44829.6				&				& 12.5				&\\
12 Oct 2016	& 2276, 2280	& B			& 200			& 23816.0				&				& 9.1					&\\
13 Oct 2016	& 2289		& A			& 200			& 40539.7				&10, 14			& 7.7					&\\
14 Oct 2016	& 2299		& B			& 200			& 37354.7				&				& 10.4				&\\
16 Oct 2016	& 2325		& A			& 200			& 44769.8 			&				& 12.4				&\\
17 Oct 2016	& 2329		& B			& 200			& 44341.9				&				& 12.3				&\\
18 Oct 2016	& 2338		& A			& 200			& 43207.2				&				& 12.0				&\\
19 Oct 2016	& 2347		& B			& 200			& 43207.2				& 				& 12.0				&\\
23 Aug 2017	& 4002		& A			& 200			& 36012.3				& 				& 8.3					&\\
27 Sep 2017	& 4399		& A			& 200			& 18005.5				&				& 5.0					&\\
\hline
\end{tabular} 
\\$\dagger$see Equation 1
\end{minipage}
\end{table*}

\section{Data processing and imaging}
\label{sec:processing}

The observations from individual beams for each night were edited, calibrated, and imaged using {\sc{ASKAPsoft}} (Whiting et al., in prep) -- the processing pipeline specifically designed and developed for ASKAP -- on the Galaxy Cray in the Pawsey Supercomputing Centre.  Processing the entirety of the Early Science data for the NGC~7232 field would have taken several weeks of computing time and resulted in over 120 TB of intermediate and final data products.  For the purpose of this paper, and in order to save time and disk space, we only used the relevant portion of the entire data set.  We processed four beams from each night (footprint A = beams 02,03,11,12;  rotated footprint A = beams 01,03,08,09; footprint B = beams 03,09,10,11; see Figure~\ref{fig:footprint}) and a reduced bandwidth of 16 MHz around the {H\sc{i}} line.  The bandwidth was split into 864 channels, maintaining the 18.5 kHz (= 3.9 km s$^{-1}$) spectral resolution, covering a frequency range between 1404.5 to 1420.5 MHz and an equivalent velocity range of -16 to 3391 km s$^{-1}$.

Most of the processing parameters were the default settings for {\sc{ASKAPsoft}} version 0.19.7\footnote{https://doi.org/10.4225/08/5ac2dc9a16430}.  These settings included the use of the automated dynamic flagging utility that identifies outlying signals as well as the use of self-calibration to correct the time dependent fluctuations of the antenna gains (both amplitude and phase) during the observation of the science target. Changes to the default processing parameters were selected to omit autocorrelations and any antennas specified in Table~\ref{table:obs}.  The calibrator observations on PKS 1934-638 were often taken at low elevation in the late afternoon.  Solar interference can introduce spurious signals that are more pronounced on the shorter baselines.  As a cautionary measure, any calibrator data from baselines shorter than 200m were excluded from the processing procedure.

In the imaging stage, due to the limited number of shorter baselines, natural weighting resulted in a ``patchy'' noise pattern across the spectral line image cube.  As such, a weighting of robust = 0.5 was chosen to better utilise the uv-coverage pattern of ASKAP-12, which resulted in smoother background noise.  The preliminary image cubes had a beam size of $\sim 22 \times 18$ arcsec.  A Gaussian taper of 30 arcsec was applied to achieve a synthesized beam of 35.5 $\times$ 30 arcsec and bring out {H\sc{i}}  of lower column density.  Due to residual artefacts from the uv-based model continuum subtraction, the spectral line image cubes for each beam also went through image-based continuum subtraction prior to being linearly mosaicked together using the {\sc{linmos}} task in {\sc{ASKAPsoft}}.  The details of the final ASKAP image cube are summarized in Table~\ref{table:image}. 

\begin{table}
 \centering
 \begin{minipage}{80mm}
 \caption{Properties of final ASKAP {H\sc{i}} image cube}
 \label{table:image}
\begin{tabular}{l c}  
\hline
On-source observing time (combined hours)		& 150 \\
Weighting	scheme							& robust = 0.5	\\
Pixel size (arcsec)							& 4 \\
Gaussian taper	(arcsec)						& 30 $\times$ 30	\\
Synthesized beam (arcsec)					& 35.5 $\times$ 30	\\
Channel width (kHz)							& 18.5 	\\
Channel width (km s$^{-1}$)					& 3.9 	\\
RMS - central region	(mJy per beam per channel)	& $\sim$2.2	\\
RMS - outer edge (mJy per beam per channel)	& $\sim$7	\\
\hline
\end{tabular} 
\end{minipage}
\end{table}

\subsection{Data Quality}

The 4-beam mosaicked image cubes for each single night of observations, which had a typical RMS noise of $\sim$7 mJy per beam per channel near the beam centres, were visually inspected and then combined to decrease the noise of the targeted region, in an attempt to reach the full WALLABY sensitivity level (see Figure~\ref{fig:rms}).  Three nights of footprint B observations (7-9 Oct 2016) showed a strong ripple pattern in the beam edge region coinciding with the triplet and were therefore not included in the final mosaic.  Overall, 150 hours (footprint A = 76h; rotated footprint A = 19h; footprint B = 55h) of equivalent ASKAP-12 observations were combined to produce the final 12-beam image cube of the NGC~7232/3 triplet and surrounding group members\footnote{this final image cube is publicly available on CASDA; https://doi.org/10.25919/5becef4d41dab}.  The highest sensitivity region of the final cube achieved an RMS noise of $\sim$2.2 mJy per beam per channel while the outer edge of the cube was up to three times noisier.

Differences between the measured and predicted RMS in Figure~\ref{fig:rms} can be attributed to the presence of faint artefacts in the image cubes.  Repeat processing with different flagging, calibration, continuum subtraction, and imaging parameters drove a significant portion of {\sc{ASKAPsoft}} development; nevertheless, due to the preliminary ASKAP-12 observing techniques as well as the early nature of the {\sc{ASKAPsoft}} processing pipeline, some artefacts remain in the final cube.  For example, residual RFI signals that were not fully excised from the data produce large scale-striping across certain channels.  There are also a handful of spatially compact negative signals that occur in channels with relatively bright {H\sc{i}} emission -- possibly due to calibration issues (see \citealt{gro2014}) or that have been introduced during the image combination phase -- which require further investigation.  

In addition, {\sc{linmos}} assumes circular Gaussian beam shapes; however, ASKAP's electronic beams are currently formed using an algorithm designed to maximise sensitivity rather than constrain beam shape (for further details on PAFs and beam forming, see \citealt{hay2008} and \citealt{hay2016}).  Holography measurements have shown that the resulting beam patterns have some ellipticity in individual polarisations and are non-Gaussian as a result of coma distortion at the edge of the field of view.  The beam shapes also vary slightly from one antenna to another.  A quantitative analysis of beam properties and behaviour will be published in future.  For this current analysis, we note that errors in the primary beam correction can contribute up to 10 percent uncertainty in source flux estimates and a higher noise level in the multi-footprint combined data (see \citealt{ser2015b}, \citealt{hey2016}, \citealt{mcc2016}). 

The aforementioned artefacts and imaging concerns do not appear to significantly affect the scientific results presented in this paper.  As more antennas are commissioned and added to the array, the increased uv-coverage will improve the sensitivity of ASKAP.  Furthermore, enhancements to the telescope (e.g.~having local ingest nodes to prevent data loss and using an on-dish calibration system to track gain/phase changes) as well as improvements in the observing technique (e.g.~using rotated footprints and shape-constrained beams) are progressively improving the data quality with each iteration of Early Science observations.

\section{Results}
\label{sec:results}

Using our final ASKAP {H\sc{i}} cube of the target region, we detect 17 {H\sc{i}} sources in the immediate vicinity of NGC~7232/3 triplet (i.e.~between 1100 - 3300 km s$^{-1}$) that are likely contained within the NGC 7232 group. Eleven of these detections are identified with stellar counterparts, of which five are newly resolved {H\sc{i}} galaxies.  The other six {H\sc{i}} detections are likely tidal debris associated with the triplet.  

Source extraction was conducted using an automated application, SoFiA (\citealt{ser2015a}).  We chose a 5-sigma threshold for source detection with SoFiA and all extracted sources were visually inspected for verification.  By choosing a high source finding threshold, most of the processing artefacts are generally rejected by SoFiA; however, not all of the {H\sc{i}} flux is recovered for each source.  Alternatively, a lower threshold detects more galaxies and more of their flux but also increases the false detection rate due to the faint artefacts in the ASKAP cube.  The number of {H\sc{i}} sources reported here is a conservative value for this field, which will be revisited during the full WALLABY survey with ASKAP-36.

\begin{figure*}
\begin{center}
  \includegraphics[width=160mm]{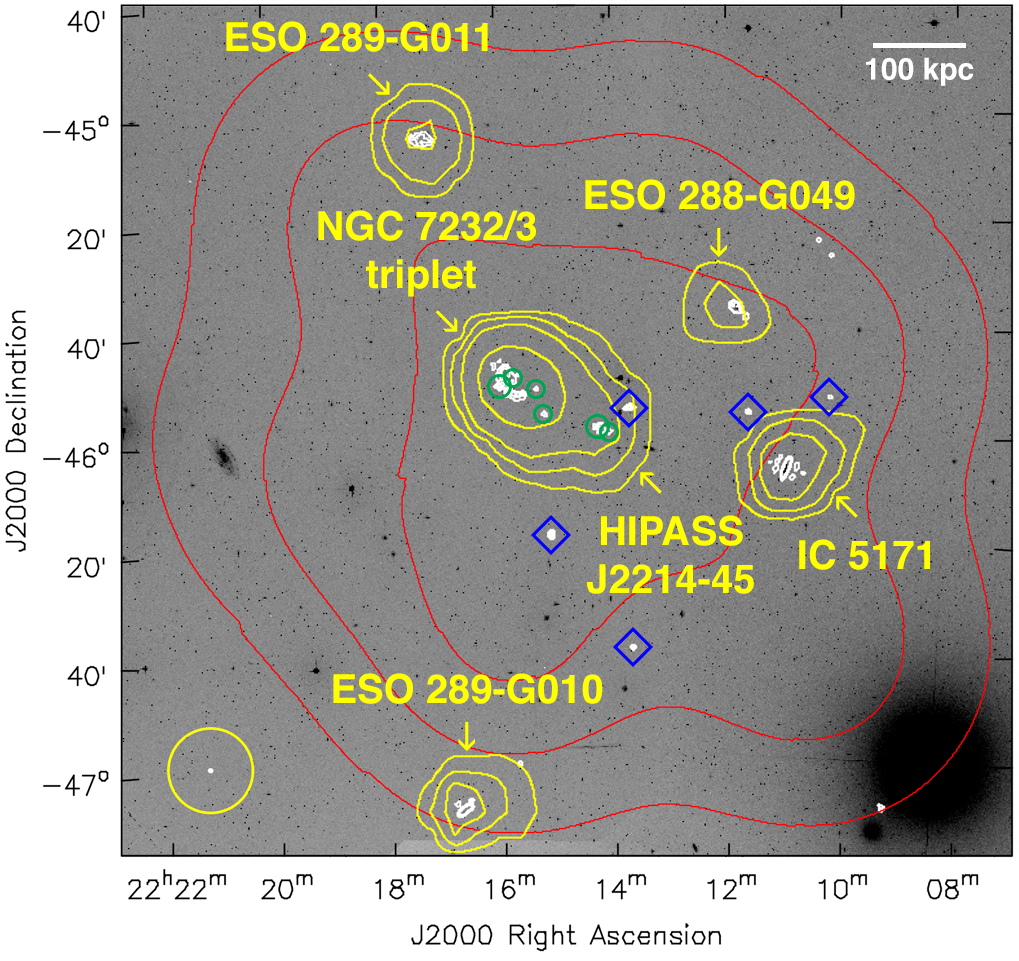}
  \caption{Total {H\sc{i}} intensity (moment 0) contours from ASKAP (white; at (1, 3, 6) $\times$ 10$^{20}$ atoms cm$^{-2}$) and HIPASS (yellow) sources within the selected velocity range -- extracted using SoFiA -- superimposed on an archival DSS2 Blue optical image.  The corresponding beam sizes for each set of {H\sc{i}} observations are shown in the bottom left and a physical scale bar based on a group distance of $\sim$24 Mpc \citep{gar1995} is in the upper right corner.  HIPASS detected sources are as labelled.  Blue diamonds indicate new ASKAP {H\sc{i}} detections that appear to have stellar counterparts, while the green circles indicate likely tidal debris from the interacting triplet, which have no detectable stellar counterparts. The red contours represent the sensitivity pattern of the 12-beam footprint, at 15, 50 and 90 percent of the peak sensitivity, based on the normalized number of visibilities used for each part of the image.  The NGC~7232/3 triplet resides in the highest sensitivity region of the imaged data.  Due to the preliminary nature of the {\sc{ASKAPsoft}} processing pipeline, a few minor artefacts remain -- including sidelobe features for IC 5171 -- were picked up by the source finder. 
\label{fig:mom0}}
\end{center}
\vspace{10cm}
\end{figure*}

For the purpose of this paper, we consider any SoFiA detections in the ASKAP cube to be authentic {H\sc{i}} sources if they coincide (spatially and spectrally) with previously detected ATCA and/or HIPASS {H\sc{i}} sources or if they appear to have a stellar counterpart.  Since we are working with ASKAP Early Science data, we assume any other detections that are identified by SoFiA but do not fit the previously mentioned criteria to be imaging artefacts or residual sidelobe features.  Table~\ref{table:cross} presents the ASKAP {H\sc{i}} sources and their spatial location, as extracted by SoFiA, crossmatched with HIPASS and ATCA detections as well as likely stellar counterparts, where applicable.  All previously known {H\sc{i}} sources that fall within the 12-beam footprint and selected velocity range were detected in the ASKAP cube.  We note that \citet{barweb2001} focus their ATCA observations on a $40 \times 40$ arcsec region centred on the NGC~7232/3 triplet.  As such, we use the wider field HIPASS cube for a more uniform comparison of all {H\sc{i}} sources detected by ASKAP.  Figure~\ref{fig:mom0} shows the SoFiA-produced {H\sc{i}} total intensity (moment 0) contours of both the ASKAP and HIPASS data.

\begin{figure*}
\begin{center}
  \includegraphics[width=180mm]{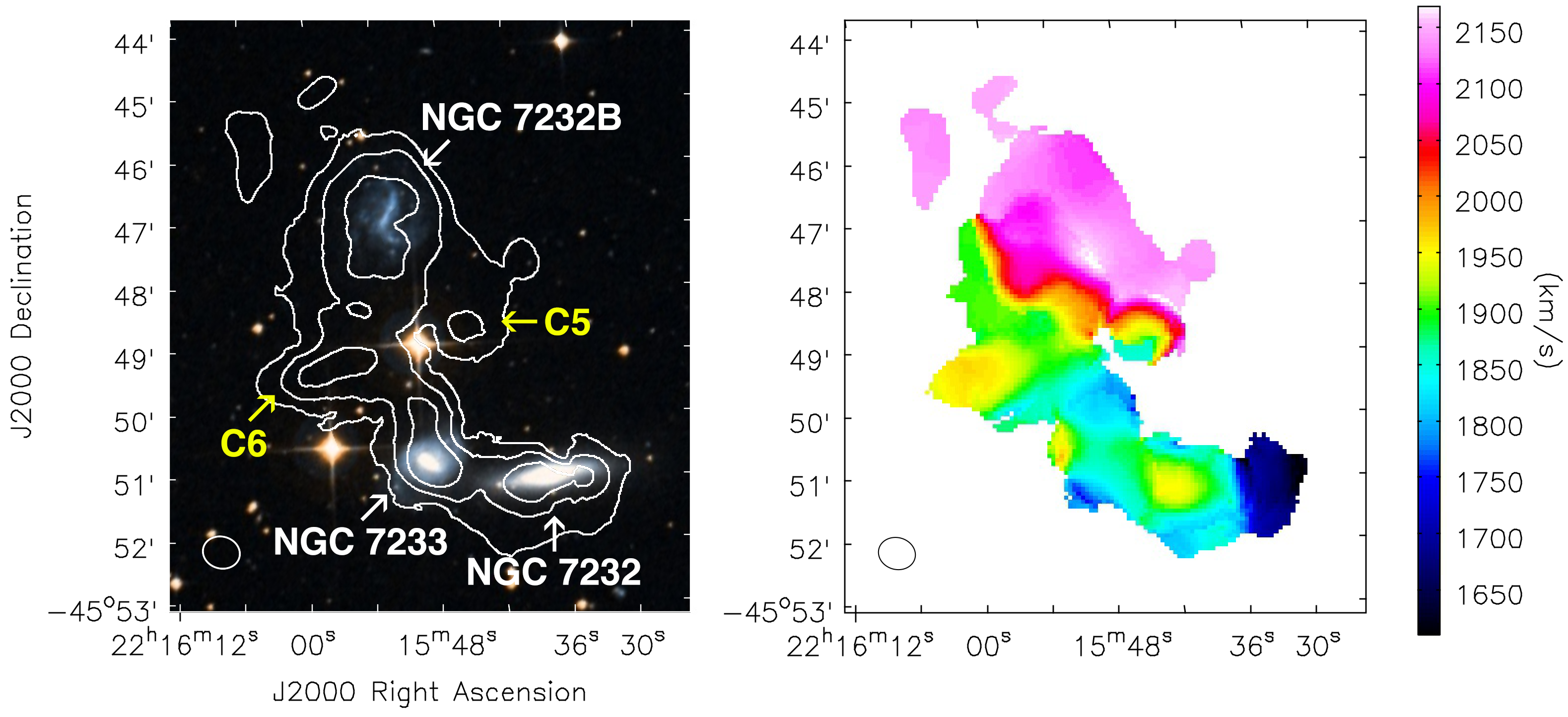}
    \caption{ASKAP {H\sc{i}} moment maps of the NGC~7232/3 triplet.  Left panel: moment 0 contours -- at (1, 3, 6) $\times$ 10$^{20}$ atoms cm$^{-2}$ -- superimposed on DSS2 coloured image.  Three {H\sc{i}} peaks coincide with the stellar components of the major galaxies.  Two additional {H\sc{i}} clouds, C5 and C6, are clearly visible in the intervening region connecting the northern spiral, NGC~7232B, to the other two galaxies NGC~7232 and NGC~7233.  Right panel: moment 1 map.  See Figure~\ref{fig:momC} for moment 1 maps with narrower velocity ranges centred around each source.
\label{fig:triplet}}
\end{center}
\end{figure*}

\begin{figure}
\begin{center}
  \includegraphics[width=80mm]{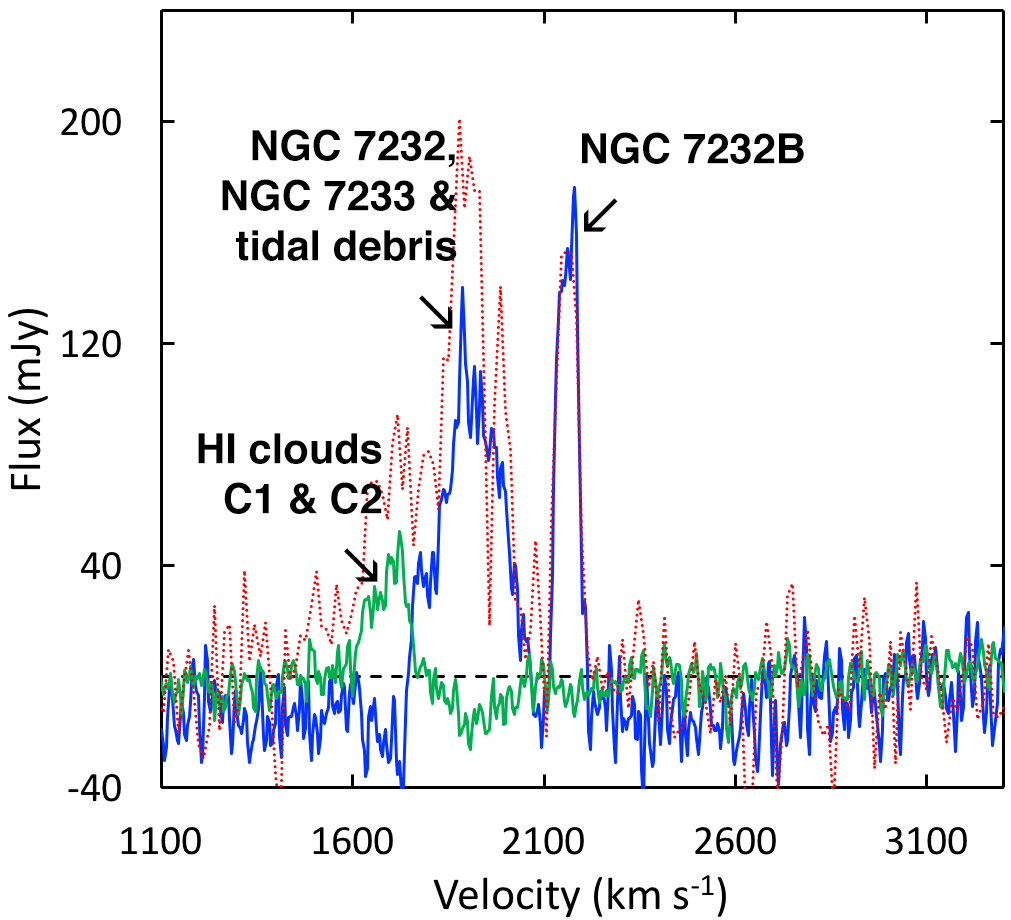}
    \caption{ASKAP {H\sc{i}} spectra of the NGC~7232/3 triplet and surrounding tidal debris.  Solid green = {H\sc{i}} clouds C1 and C2 (extracted from the ASKAP cube), solid blue = ASKAP, red dotted = HIPASS.  Due to the preliminary nature of the v.0.19.7 processing pipeline, over-subtraction of the continuum near bright {H\sc{i}} sources has resulted in negative features in the extracted spectra.
\label{fig:spectra}}
\end{center}
\end{figure}

The {H\sc{i}} emission from the members directly associated with the interacting triplet system, shown in Figure~\ref{fig:triplet}, is complex.  In the spectra presented in Figure~\ref{fig:spectra}, NGC 7232B is clearly differentiated by its unique velocity range (2120 - 2230 km s$^{-1}$) in both the ASKAP and HIPASS observations.  ASKAP provides sufficient angular resolution to distinguish six {H\sc{i}} clouds -- referred to as {H\sc{i}} clouds C1-C6 -- from the main body of the triplet (see Figure~\ref{fig:momC}). 

Individual moment maps for {H\sc{i}} sources in the surrounding area can be found in the appendix.  {H\sc{i}} galaxies that are also detected in HIPASS and are fairly well resolved in the ASKAP data are shown in Figure~A\ref{fig:momA}.  Newly detected ASKAP {H\sc{i}} sources that appear to have stellar counterparts (indicated by the blue diamonds in Figure~\ref{fig:mom0}) are shown in Figure~A\ref{fig:momB}.  All {H\sc{i}} velocity (moment 1) maps were generated by applying the SoFiA output mask to the original data cube.  Additional masking has been applied for the moment 1 maps of {H\sc{i}} clouds C1-C6 in Figure~\ref{fig:momC}.  For each ASKAP detected {H\sc{i}} source, manually defined ellipses on the unmasked cube -- using the SoFiA outputs as a guide -- were used to create all spectral profiles.  

\begin{figure*}
\begin{center}
  \includegraphics[width=180mm]{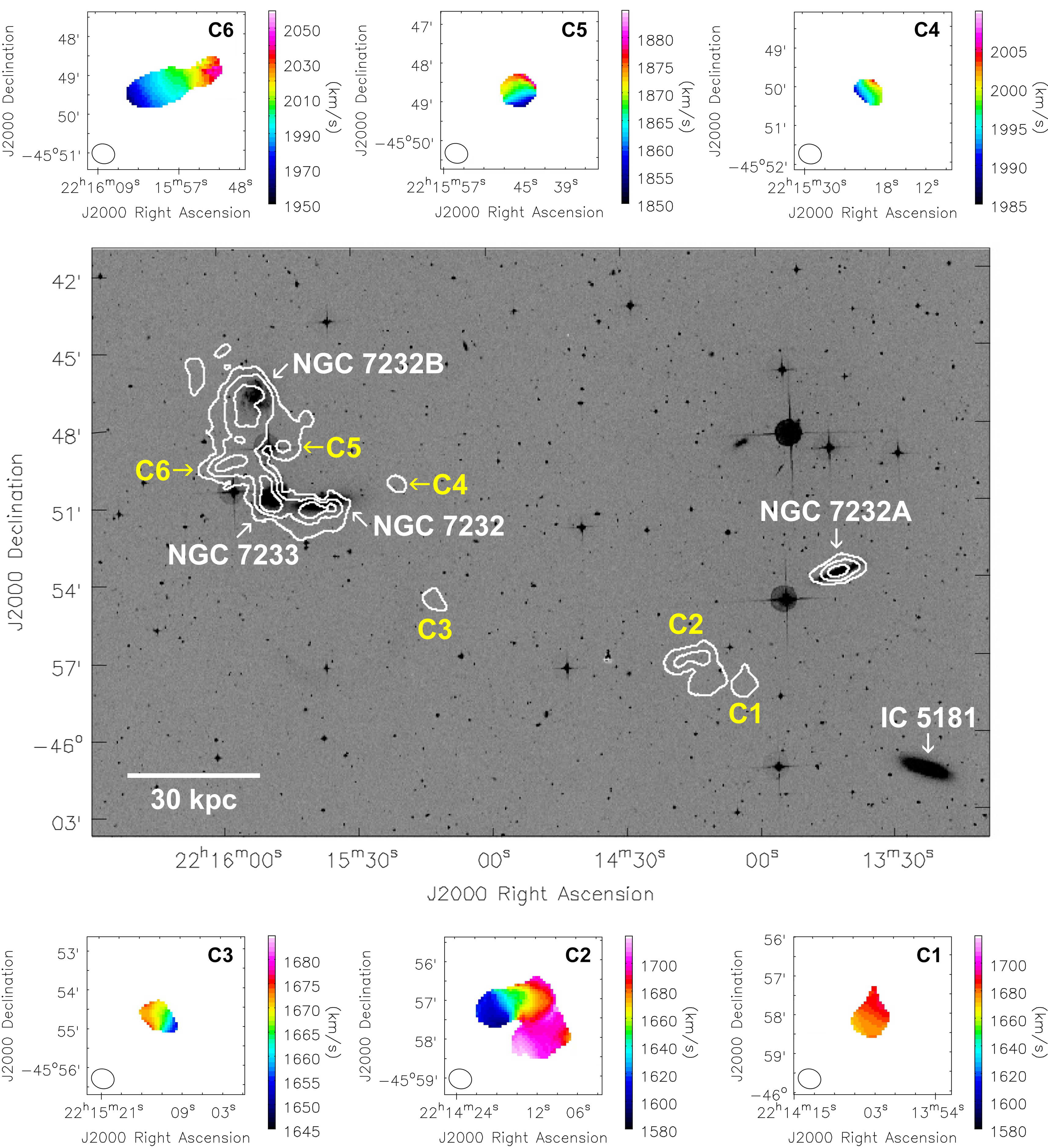}
    \caption{ASKAP {H\sc{i}} emission associated with the NGC~7232/3 triplet.   Centre: ASKAP {H\sc{i}} moment 0 contours -- at (1, 3, 6) $\times$ 10$^{20}$ atoms cm$^{-2}$ -- superimposed on DSS2 Blue archival images.  Pertinent optical galaxies are labelled in white.  A physical scale bar assuming a group distance of 24 Mpc is shown in the bottom left of the panel.  Outer: ASKAP moment 1 maps of the {H\sc{i}} emission in the velocity range centred around each source, as labelled.  Additional masking of neighbouring sources has been applied to {H\sc{i}} clouds C1-C6.  On accounts of their spatial and spectral proximity, {H\sc{i}} cloud C1 and C2 are shown with the same velocity colour scale.
\label{fig:momC}}
\end{center}
\end{figure*} 

Table~\ref{table:summary} summarizes the {H\sc{i}} properties of each source, as measured by SoFiA.  The uncertainty in the central velocity (v$_{\mathrm{HI}}$) is
\begin{equation} 
\sigma(\mbox{v}_{\mathrm{HI}}) = 3(S/N)^{-1}(P \delta v)^{1/2},
\end{equation}
where $S/N$ is the median signal-to-noise ratio of the flux for each source, $P$ is half the difference of the velocity width measured at 50 percent (W$_{50}$) and 20 percent (W$_{20}$) of the peak flux and $\delta v$ = 3.9 km $s^{-1}$ is the spectral resolution of the ASKAP data (see \citealt{kor2004} for derivation details).  Uncertainties in the line widths are $\sigma$(W$_{50}$) = 2$\sigma$(v$_{\mathrm{HI}}$) and $\sigma$(W$_{20}$) = 3$\sigma$(v$_{\mathrm{HI}}$).  The 20 percent flux uncertainty takes into account calibration/processing as well as instrumental effects.  Even with the 5-sigma detection threshold, the SoFiA computed fluxes are, on average, within 15 percent of the {H\sc{i}} flux manually computed from the spectral profiles of each galaxy.  Table~\ref{table:summary} also includes the local RMS noise level around each source and the integrated S/N of the detection.  A common distance of 24 Mpc is assumed for all group members to compute the {H\sc{i}} masses (M$_{\mathrm{HI}}$).

\begin{landscape}
\begin{table}
 \caption{Crossmatches of the ASKAP {H\sc{i}} sources detected in the NGC~7232 group} 
 \label{table:cross}
\begin{tabular}{ l c c c c c c @{} c @{} c}  
\hline
Source Name	 			&ASKAP RA, Dec	&HIPASS 		&HIPASS v$_{\mathrm{HI}}$ $^1$	& HIPASS flux$^1$ 		&ATCA v$_{\mathrm{HI}}$ $^2$	&ATCA flux$^2$	 &Stellar counterpart			&v$_{\mbox{stellar}}$\\
						&(J2000)			&designation	&(km s$^{-1}$)					& (Jy km s$^{-1}$)		&(km s$^{-1}$)		& (Jy km s$^{-1}$)	&						&(km s$^{-1}$)\\
\hline 
WALLABY J221010-455159	& 22:10:10.30, -45:51:59.4	& --			& --				& --				& --				& --				& MRSS 288-021830			& --\\
WALLABY J221056-460500	& 22:10:56.45, -46:05:00.8	& J2210-46	& 2990.9			& 15.8			& --				& --				& IC 5171						& 2827 $\pm$ 26 $^3$\\
WALLABY J221136-455439	& 22:11:36.26, -45:54:39.6	& --			& --				& --				& --				& --				& AM 2208-460					& --\\
WALLABY J221148-453525	& 22:11:48.88, -45:35:25.9	& J2211-45	& 1918.2			& 6.6				& --				& --				& ESO 288-G049				& 1968 $^4$\\
WALLABY J221339-463746	& 22:13:39.98, -46:37:46.9	& --			& --				& --				& --				& --				& MRSS 289-168885			& --\\
WALLABY J221341-455343 	& 22:13:41.60, -45:53:43.3	& -- 			& --				& --				& --				& --				& NGC 7232A (ESO 289-G003) 	& 2224 $\pm$ 55 $^5$\\
WALLABY J221403-455806	& 22:14:03.51, -45:58:06.9	& -- 			& --				& --				& --				& --				& --							& --\\ 
WALLABY J221413-455723	& 22:14:13.86, -45:57:23.1	& J2214-45	& 1864.8			& 11.9			& --				& --				& --							& -- \\ 
WALLABY J221506-461658	& 22:15:06.41, -46:16:58.4	& --			& --				& --				& --				& --				& ESO 289-G005				& 1885 $\pm$ 55 $^5$\\
WALLABY J221512-455442 	&22:15:12.23, -45:54:42.5		& -- 			& --				& --				& --				& --				& --							& --\\ 	
WALLABY J221520-455009 	&22:15:20.28, -45:50:09.6		& -- 			& --				& --				& --				& --				& --							& --\\ 	
WALLABY J221545-454843 	&22:15:45.79, -45:48:43.3		& -- 			& --				& --				& --				& --				& --							& --\\ 	
WALLABY J221547-455007	& 22:15:47.67, -45:50:07.0	& J2215-45a	& 1904.9			& 24.6			& 1735$^\dagger$			& 3.5$^\dagger$				& NGC 7232 and NGC 7233 			& 1887 $\pm$ 25 $^3$, 1862 $\pm$ 26 $^3$\\ 
WALLABY J221551-454702	& 22:15:51.58, -45:47:02.7	& J2215-45b	& 2158.9			& 9.5				& 2160			& 7.2				& NGC 7232B (ESO 289-G009) 	& 2152 $\pm$ 45 $^5$\\
WALLABY J221556-454903	& 22:15:56.97, -45:49:03.9	& -- 			& --				& --				& --				& --				& --							& --\\ 
WALLABY J221642-470708	& 22:16:42.68, -47:07:08.5	& J2216-47	& 2681.7			& 12.2 			& --				& --				& ESO 289-G010				& 2785 $^4$\\ 
WALLABY J221716-450359	& 22:17:16.45, -45:03:59.4	& J2217-45a	& 1784.7			& 10.5			& --				& --				& ESO 289-G011				& 1822 $\pm$ 20  $^3$\\
\hline
\end{tabular} 
\\$^1$\citet{mey2004}, $^2$\citet{barweb2001}, $^3$\citet{dac1991}, $^4$\citet{tul2008}, $^5$\citet{jon2009}.  $^\dagger$These values are for NGC~7232 only as NGC~7233 was reported as an unresolved source by \citet{barweb2001}, see Section~\ref{sec:discuss} for more details.
\end{table}
\end{landscape}

\begin{landscape}
\begin{table}
 \caption{ASKAP {H\sc{i}} properties of the sources detected in the NGC~7232 group}
 \label{table:summary}
\begin{tabular}{ l c c c c c c c c}  
\hline
Source name	 	& ASKAP v$_{\mathrm{HI}}$	& ASKAP W$_{50}$	& ASKAP W$_{20}$	& ASKAP flux		&Local RMS		&Integrated		& ASKAP M$_{\mathrm{HI}}$	&Comments\\
						& (km s$^{-1}$)		& (km s$^{-1}$)		& (km s$^{-1}$)		& (Jy km s$^{-1}$)	&(mJy beam$^{-1}$)	&S/N		&($\times 10^8$ M$_{\odot}$)	&\\
\hline 
WALLABY J221010-455159	&2026 $\pm$ 3		&44 $\pm$ 6		&50 $\pm$ 9		&0.3 $\pm$ 0.1		&2.2				&6		&0.5 $\pm$ 0.1				& MRSS 288-021830, first {H\sc{i}} detection\\
WALLABY J221056-460500	&2831 $\pm$ 3		&365 $\pm$ 6		&380 $\pm$ 10		&11 $\pm$ 2		&2.5				&49		&14 $\pm$ 3				& IC 5171\\
WALLABY J221136-455439	&1953 $\pm$ 6		&30 $\pm$ 10		&40 $\pm$ 20		&0.6 $\pm$ 0.1		&2.2				&10		&0.8 $\pm$ 0.2				& AM 2208-460	, first {H\sc{i}} detection\\
WALLABY J221148-453525	&1964 $\pm$ 2		&155 $\pm$ 4		&168 $\pm$ 7		&6 $\pm$ 1		&2.4				&38		&8 $\pm$ 2				& ESO 288-G049\\
WALLABY J221339-463746	&1741 $\pm$ 3		&46 $\pm$ 5		&50 $\pm$ 8		&0.5 $\pm$ 0.1		&2.5				&7		&0.6 $\pm$ 0.1				& MRSS 289-168885, first {H\sc{i}} detection\\
WALLABY J221341-455343 	&2348 $\pm$ 5		&220 $\pm$ 10		&230 $\pm$ 10		&2.0 $\pm$ 0.4		&2.3				&15		&2.7 $\pm$ 0.5				& NGC 7232A, first {H\sc{i}} detection\\
WALLABY J221403-455806	&1692 $\pm$ 6		&40 $\pm$ 10		&60 $\pm$ 20		&0.7 $\pm$ 0.1		&2.4				&7		&1.0 $\pm$ 0.2				& {H\sc{i}} cloud C1\\
WALLABY J221413-455723	&1678 $\pm$ 5		&117 $\pm$ 9		&150 $\pm$ 10		&2.4 $\pm$ 0.5		&2.4				&15		&3.3 $\pm$ 0.7				& {H\sc{i}} cloud C2\\ 
WALLABY J221506-461658	&1931 $\pm$ 7		&130 $\pm$ 10		&150 $\pm$ 20		&2.1 $\pm$ 0.4		&2.4				&18		&2.9 $\pm$ 0.6				& ESO 289-G005, first {H\sc{i}} detection\\
WALLABY J221512-455442	&1668 $\pm$ 5		&30 $\pm$ 10		&50 $\pm$ 20		&0.4 $\pm$ 0.1		&2.4				&6		&0.6 $\pm$ 0.1				& {H\sc{i}} cloud C3\\
WALLABY J221520-455009	&1999 $\pm$ 5		&30 $\pm$ 10		&40 $\pm$ 20		&0.3 $\pm$ 0.1		&2.3				&5		&0.4 $\pm$ 0.1				& {H\sc{i}} cloud C4\\
WALLABY J221545-454843	&1874 $\pm$ 6		&30 $\pm$ 10		&60 $\pm$ 20		&0.5 $\pm$ 0.1		&2.2				&6		&0.7 $\pm$ 0.1				& {H\sc{i}} cloud C5\\
WALLABY J221547-455007	&1863 $\pm$ 4		&152 $\pm$ 7		&260 $\pm$ 10		&12 $\pm$ 2		&2.3				&33		&17 $\pm$ 3				& NGC~7232 and NGC 7233 \\ 
WALLABY J221551-454702	&2159 $\pm$ 1		&64 $\pm$ 2		&81 $\pm$ 3		&7 $\pm$ 1		&2.3				&38		&9 $\pm$ 2				& NGC 7232B\\ 
WALLABY J221556-454903	&2002 $\pm$ 7		&40 $\pm$ 10		&110 $\pm$ 20		&2.1 $\pm$ 0.4		&2.4				&15		&2.8 $\pm$ 0.6				& {H\sc{i}} cloud C6\\ 
WALLABY J221642-470708	&2768 $\pm$ 4		&230 $\pm$ 8		&240 $\pm$ 10		&12 $\pm$ 2		&4.8				&32		&16 $\pm$ 3				& ESO 289-G010 \\ 
WALLABY J221716-450359	&1821 $\pm$ 1		&113 $\pm$ 2		&121 $\pm$ 3		&8 $\pm$ 2		&3.4				&30		&10 $\pm$ 2				& ESO 289-G011\\ 
\hline
\end{tabular} 
\end{table}
\end{landscape} 

\subsection{Stellar properties of the NGC~7232/3 triplet}

To further explore and possibly disentangle some of the complexity of the triplet system, we used ancillary data to estimate the stellar properties of the three major optical galaxies, NGC~7232, NGC~7232B and NGC~7233.  We measured the $g$- and $r$-band fluxes using SkyMapper images \citep{wol2018}.  These fluxes were then used estimate stellar masses (M$_{*}$) by applying the relation from \citet{bel2003}.  We measured the far-ultraviolet (FUV) flux from GALEX \citep{mar2005} and W4 (22~$\mu$m) flux from WISE \citep{wri2010} to estimate the star formation rate (SFR) following the method described in \citet{wan2017}.  SFR$_{\mbox{W4}}$ indicates the dust-attenuated SFR and SFR$_{\mbox{FUV}}$ is the unattenuated portion.  As such, the sum of these two values provides the total SFR \citep{wan2017}.  All stellar flux measurements were obtained using Petrosian apertures and with neighbouring galaxies masked out of the images.  The $r$-band image was used as the reference for setting the apertures on the SkyMapper images.  For GALEX and WISE, the band which produced the largest Petrosian ellipse determined the final aperture used for each galaxy. The stellar properties for the galaxies within the triplet are presented in Table~\ref{table:stellar}. 

\begin{table*}
 \centering
 \begin{minipage}{150mm}
 \caption{Stellar properties of the NGC~7232/3 triplet}
 \label{table:stellar}
\begin{tabular}{ l c c c c c c c c}  
\hline
Source name	&$r$-band			&$g$-band		&M$_{\mbox{*}}$		&FUV				&SFR$_{\mbox{FUV}}$	&W4				&SFR$_{\mbox{W4}}$	&SFR$_{\mbox{total}}$\\
			&magnitude$^1$	&magnitude$^1$	&$\times 10^8$ M$_{\odot}$	&magnitude$^2$		&M$_{\odot}$ yr$^{-1}$	&magnitude$^3$	&M$_{\odot}$ yr$^{-1}$	&M$_{\odot}$ yr$^{-1}$\\
\hline 
NGC~7232	&11.99 $\pm$ 0.03	& 12.65 $\pm$ 0.03 		&120					& 16.93 $\pm$ 0.06 		& 0.04				& 6.24 $\pm$ 0.03	& 0.05				&0.09\\
NGC~7232B	&12.65 $\pm$ 0.03	& 13.24 $\pm$ 0.03		&55					& 15.73 $\pm$ 0.05		& 0.12				& 2.70 $\pm$ 0.01	& 1.3					&1.4\\
NGC~7233	&15.7 $\pm$ 0.2	& 15.9$\pm$ 0.1		&1.0					& 14.67 $\pm$ 0.05		& 0.3					& 9.1 $\pm$ 0.2	& 0.004				&0.3\\
\hline
\end{tabular} 
values measured from $^1$SkyMapper \citep{wol2018}, $^2$GALEX \citep{mar2005}, $^3$WISE \citep{wri2010}
\end{minipage}
\end{table*}
\vspace{10mm} 

\section{Analysis and discussion}
\label{sec:discuss}

The capabilities of ASKAP, during its Early Science phase, have enabled us to detect and/or resolve the {H\sc{i}} emission of several galaxies.  Within the central 12-beam footprint targeting the NGC~7232 group, there are 17 {H\sc{i}} sources.  Many of these sources appear to be star-forming galaxies with clear optical counterparts and are further detailed in the appendix.  This section will discussion the individual sources within the NGC~7232/3 triplet and its neighbouring {H\sc{i}} clouds.

Due to the proximity of NGC~7232, NGC~7232B and NGC~7233 as well as the common {H\sc{i}} envelope detected by HIPASS (see Figure~\ref{fig:mom0}), it appears that these galaxies are actively interacting (\citealt{barweb2001}, \citealt{kor2004}).  The {H\sc{i}} corresponding to the optical galaxies within the triplet is noticeably disturbed.  There appears to be an {H\sc{i}} bridge connecting all three galaxies (Figure~\ref{fig:triplet}) as well as six tidally formed {H\sc{i}} clouds in the neighbouring area (Figure~\ref{fig:momC}).  One of these clouds, {H\sc{i}} cloud C6, was originally detected as an {H\sc{i}} plume by \citet{barweb2001} in their ATCA observations and has sufficient mass (i.e.~M$_{\mathrm{HI}} > 10^8$ M$_{\odot}$) to become a self-gravitating TDG (\citealt{lel2015}, \citealt{lee2016}).  The other clouds are likely more transient tidal features that will fade into the lower column density {H\sc{i}} envelope or fall into one of the larger galaxies (see \citealt{bou2006}).

Adding together the HIPASS values for the NGC 7232/3 triplet and neighbouring gas cloud HIPASS J2214-45 (reported by \citet{mey2004}; see Table~\ref{table:cross}), the total {H\sc{i}} flux for this region is 46 Jy km s$^{-1}$; however, HIPASS J2214-45 is likely a confused source that includes some {H\sc{i}} emission also attributed to the NGC~7232 and NGC~7233 pair (i.e. HIPASS J2214-45a).  \citet{kor2004}, cataloguing the 1000 brightest sources in HIPASS, report an {H\sc{i}} flux of 34.6 $\pm$ 4.1 Jy km s$^{-1}$ for this same region (identified as the AM2212-460 group). From the ASKAP data, we measure a total {H\sc{i}} flux of 26 $\pm$ 3 Jy km s$^{-1}$ for the triplet and surrounding {H\sc{i}} clouds.  Looking at the spectra in Figure~\ref{fig:spectra}, ASKAP appears to recover at least 70 percent of the {H\sc{i}} flux in the spatial and spectral vicinity of the NGC 7232/3 triplet, which is in agreement with the HIPASS value reported by \citet{kor2004}.

The $\sim$25 percent flux difference between ASKAP and HIPASS can be attributed to the column density sensitivity limit of the ASKAP data as well as the high source detection threshold chosen for SoFiA.  Another minor factor with be the slight negative offset in the current ASKAP spectra (see Figure~\ref{fig:spectra}) due to over-subtraction of the continuum near bright {H\sc{i}} sources.  Since the galaxies within this region are interacting, it is also likely that a portion of the diffuse gas is resolved out by the interferometer.

\subsubsection*{NGC~7232 and NGC 7233}

The {H\sc{i}} components of NGC~7232 and NGC~7233 are thoroughly blended together and span a broad velocity range (see Figures~\ref{fig:triplet} and \ref{fig:spectra} ).  There does appear to be higher density clumps of {H\sc{i}} coinciding with the stellar disks of the two galaxies; nevertheless, much of the {H\sc{i}} from these galaxies appears to be distributed throughout the spatial region of the triplet system. Discrepancies between source detection and separation has resulted in the inconsistencies of the {H\sc{i}} properties, reported in Tables~\ref{table:cross} and \ref{table:summary}, from each set of observations on NGC~7232 and NGC 7233.

The wider variety of array baseline lengths and longer integration time of the ASKAP observations enable greater sensitivity to extended emission than the ATCA observations by \citet{barweb2001}.  Although the ATCA observations detect NGC~7233 (see the ATCA {H\sc{i}} moment 0 map in figure 8 of \citealt{barweb2001}), it was determine by those authors that the source is unresolved and no further {H\sc{i}} measurements were reported. Comparing the {H\sc{i}} moment maps, ASKAP does appear to recover more {H\sc{i}} emission that ATCA; however, about half of the {H\sc{i}} flux that was catalogued by HIPASS has not been recovered by ASKAP.  
Subtracting the ASKAP detections from the HIPASS cube indicates that there is diffuse gas in the area between the NGC~7232/3 triplet and a neighbouring galaxy, IC~5181, suggesting that there is an {H\sc{i}} tail/bridge in this region. Note, no {H\sc{i}} counterpart has been detected across the stellar extent of IC~5181(which is a lenticular galaxy located at RA = 22:13:22, Dec = -46:01:03, v$_{\mbox{stellar}} = 1936 \pm 45$ km s$^{-1}$; \citealt{jon2009}).

The stellar component of NGC~7232 is ten times more massive than that of NGC~7233 (Table~\ref{table:stellar}).  Although we are unable to distinguish the {H\sc{i}} originating from each galaxy, since they are located within the same group environment, one can assume that NGC~7232 would have started with more {H\sc{i}} than NGC~7233 (see \citealt{den2014} for more details on mass scaling relations).  Currently, the {H\sc{i}} peaks coinciding with the optical centres of the two galaxies are quite comparable in size and column density, suggesting that most of the {H\sc{i}} originating from NGC~7232 is now spread throughout triplet system, which may explain this galaxy's low SFR.

\subsubsection*{NGC 7232B}

The {H\sc{i}} associated with NGC~7232B, a face-on spiral, is centred at v$_{\mathrm{HI}} = 2159 \pm 1$ km s$^{-1}$ and is kinematically distinct from NGC ~7232 and NGC~7233, which span a {H\sc{i}} velocity range of v$_{\mathrm{HI}} \sim$1600 - 2100 km s$^{-1}$ (see Figure~\ref{fig:spectra}).  The gas in the outer region of NGC~7232B shows significant signs of spatial distortion.  In the moment maps of this galaxy (Figure~\ref{fig:triplet}), there appears to be a southwest extension towards {H\sc{i}} cloud C5 and a clump of gas east of the central region of the galaxy.  ASKAP recovers the same amount of {H\sc{i}} flux as ATCA and over 70 percent of the emission detected by HIPASS for this galaxy.  This result suggests that most of the gas associated with NGC~7232B is not too diffusely spread out and easily recovered by the two interferometers.

NGC~7232B is actively star forming, especially compared to the other two galaxies in the system.  The {H\sc{i}} mass fraction (M$_{\mathrm{HI}}$/M$_{*}$) for NGC~7232B (Tables~\ref{table:summary} and \ref{table:stellar}) is comparable to other gaseous galaxies of similar mass in the local universe \citep{wan2017}.  This result indicates that the {H\sc{i}} associated with NGC~7232B remains primarily with this galaxy and has yet to be significantly affected (i.e.~tidally stripped) by the ongoing interaction event within the triplet system.    

\subsubsection*{{H\sc{i}} clouds C1 and C2}

Interactions between gas-rich galaxies can produce tidal tails several 100 kpc in length (e.g.~\citealt{lei2016}, \citealt{oos2018}) and within these tails, high density clumps of {H\sc{i}} are formed.  Under the right circumstances, these {H\sc{i}} clumps can accrete sufficient amounts of material to eventually become self-gravitating TDGs (\citealt{mir1992}, \citealt{lel2015}, \citealt{lee2016}).   {H\sc{i}} clouds C1 and C2 are located in the densest region of HIPASS J2214-45, the latter being a confused HIPASS source with NGC~7232 and NGC~7233.  The high-resolution ASKAP observations recover $\sim$30 percent of the {H\sc{i}} attributed to HIPASS J2214-45.  The remaining portion of the {H\sc{i}} gas is likely fairly diffuse and resides in the aforementioned tidal bridge connecting the NGC~7232/3 triplet with other group members.

{H\sc{i}} cloud C1 has a relatively low {H\sc{i}} mass and is more likely to be a transient tidal feature as it just reaches the $10^8$ M$_{\odot}$ mass threshold for long term survivability \citep{bou2006}; whereas, {H\sc{i}} cloud C2 has an {H\sc{i}} mass of $3.3 \pm 0.7 \times 10^{8}$ M$_{\odot}$.  In the moment 1 map of Figure~\ref{fig:momC}, the northeast portion of {H\sc{i}} cloud C2 appears to have a smooth velocity gradient.  If this portion of the cloud is rotating, then its estimated total dynamical mass would be on the order of M$_{\mbox{dyn}} \sim 10^9$ M$_{\odot}$ (see equation 2 in \citealt{lee2016}).  However, the lack of a clear stellar counterpart combined with this potentially large dynamical mass measurement implies that the velocity gradient across the northeast portion of {H\sc{i}} cloud C2 is not due to self-gravitation induced rotation.  Rather, it may be the result of shearing effects from the interaction process between the larger group members.  This mass discrepancy between the baryonic (i.e.~gas and stellar) mass and total mass does not take into account the ionized gas component, which can be up to three times the mass of the {H\sc{i}} gas in tidal features (see \citealt{fox2014}).  Nevertheless, the irregular morphology of the southwest portion of {H\sc{i}} cloud C2 also opposes the rotation scenario.  With the current data, we are unable to assess the longevity of {H\sc{i}} cloud C2.

It appears that IC~5181 -- located $\sim$150 kpc in projection from the NGC~7232/3 triplet, at the distance of the group -- forms a line with {H\sc{i}} clouds C1 and C2 and the triplet (see Figure~\ref{fig:momC}).  Previous optical observations of IC~5181 by \citet{piz2013} suggest that the ionized gas component along the polar regions of this galaxy have an external origin.  These {H\sc{i}} clouds could be remnants of the accretion / merging event that deposited ionized gas onto the undisturbed stellar disk of IC~5181 \citep{piz2013}, or tidal debris from the NGC~7232/3 interaction -- if the events are separate.

\subsubsection*{{H\sc{i}} clouds C3 and C4}

{H\sc{i}} clouds C3 and C4 are likely high density peaks associated with the tidal bridge.  Although these clouds are low in {H\sc{i}} mass and likely to be short-lived \citep{bou2006}, they show velocity gradients that are a testament to the high spatial and spectral capabilities of ASKAP.  Looking closely at the ASKAP data, there appears to be a tentative detection of very faint {H\sc{i}} emission connecting {H\sc{i}} cloud C3 to the triplet. However, given that much of the gas in the vicinity of {H\sc{i}} clouds C3 and C4 is diffusely distributed and/or below our detection threshold, it is difficult to discern the exact origin and significance of these tidal debris sources.

\subsubsection*{{H\sc{i}} cloud C5}

{H\sc{i}} cloud C5 spatially coincides with the southwest extension from NGC~7232B; however, the former has a lower velocity range that is similar to the {H\sc{i}} emission associated with NGC~7232 and NGC~7233.  Within the image cube {H\sc{i}} cloud C5 emerges as a distinct object, brightest between 1870 - 1890 km s$^{-1}$, alongside the other {H\sc{i}} clouds associated with the major members of the triplet.  Based on its location, {H\sc{i}} cloud C5 could be part of a gaseous tidal bridge that is forming between NGC~7232B and the other two spirals in the triplet.

\subsubsection*{{H\sc{i}} cloud C6}

{H\sc{i}} cloud C6 is quite bright and fairly compact with its peak flux at $\sim$1990 km s$^{-1}$.  Currently, it appears to be spatially and spectrally embedded in the gas of its parent galaxies but this source has enough mass to become self-gravitating \citep{bou2006}.  The smooth velocity gradient in Figure~\ref{fig:momC} suggests rotation, but could be the result of shearing as {H\sc{i}} cloud C6 is moving away from the triplet system.  Assuming the former scenario and that {H\sc{i}} cloud C6 is in dynamical equilibrium, then estimated total dynamical mass for this object would be on the order of M$_{\mbox{dyn}} \sim 10^8$ M$_{\odot}$.  This dynamical mass value is comparable to M$_{\mathrm{HI}}$, which makes rotation a plausible explanation and would indicate that {H\sc{i}} cloud C6 could be the progenitor of a dark matter poor TDG (see \citealt{lel2015}, \citealt{lee2016}).

Although the triplet system is complex -- and further complicated by projection effects -- it is interesting that {H\sc{i}} clouds C5 and C6 are aligned almost perpendicular to a line connecting NGC~7232B and NGC~7233 (Figure~\ref{fig:triplet}).  It is possible that these clouds indicate two tails that are currently forming from the interaction between the two face-on spirals in the triplet (\citealt{too1972}, \citealt{bou2006}).

With its high column density ($>6 \times 10^{20}$ atoms cm$^{-2}$ in the central region), {H\sc{i}} cloud C6 has a sufficient {H\sc{i}} gas density for star formation (see \citealt{sch2004} for details about star formation thresholds).  There appears to be no stellar over-densities associated with this cloud; nevertheless, the archival DSS2 images are likely too shallow to detect the low-surface brightness counterparts that are associated with some tidally-formed {H\sc{i}} features (e.g.~\citealt{lee2014}, \citealt{jan2015}, \citealt{mad2018}). Deeper optical imaging could possibly detect faint stellar components related to initial star formation; however, two bright (B-band magnitude $\sim$10) foreground stars HD 211111 (at RA = 22:15:50, Dec = -45:48:56) and HD 211121 (at RA = 22:15:58, Dec = -45:50:35) could hinder follow-up observations.

\section{Conclusion}
\label{sec:conclude}

The ASKAP Early Science dataset presented in this paper verifies that the array and its associated processing pipeline {\sc{ASKAPsoft}} -- both in moderately preliminary states -- are successfully producing scientifically useful data.  Overall, within the 12-beam image cube centred on the NGC~7232 group, we detect 17 {H\sc{i}} sources.  Six of these detections are well-known {H\sc{i}}-rich galaxies including one fully interacting pair.  Five of these detection are newly resolved {H\sc{i}} galaxies with identifiable stellar counterparts.  The remaining six {H\sc{i}} detections are likely tidal debris associated with the NGC~7232/3 triplet.  

The triplet is a complex system.  The {H\sc{i}} components of NGC~7232 and NGC~7233 appear to be fully intertwined. NGC~7232B still retains most of its gas but shows evidence that it is beginning to interact with the other two galaxies.  {H\sc{i}} clouds C5 and C6 are likely tidal clumps that have been produced by the triplet system and possibly indicate the projected spatial location of two tidally-formed tails of {H\sc{i}}.  If {H\sc{i}} cloud C6 is moving away from its parent galaxies, it has sufficient mass to eventually decouple from the tidal tail and possibly develop into a long-lived TDG. 

{H\sc{i}} clouds C1-C4 might have been produced by the triplet system, or these clouds could indicate an earlier interaction event involving another group member, such as IC~5181.  A portion of the {H\sc{i}} within this region, which was originally identified as belonging to HIPASS J2214-45 is likely diffusely distributed and lies below the detection threshold of our current ASKAP observations.  Accordingly, it is difficult to ascertain the origins of the {H\sc{i}} bridge that appears to be connecting NGC~7232/3 and IC~5181.  Future work would include adding short-spacing / single-dish observations to the ASKAP data in order to recover more of the diffuse gas in and around the tidal features, in order to further understand the interaction processes that are taking place.

The high-resolution capabilities of ASKAP produced moderately well-resolved moment maps of four galaxies (i.e. IC~5171, ESO~288-G049, ESO~298-G010 and ESO~289-G011) that can be used for further kinematic analysis.  The newly detected {H\sc{i}} counterparts of five stellar galaxies indicate their membership to the group.  These nine galaxies provide a more complete picture of the group environment surrounding the NGC~7232/3 triplet.

There are many areas where the data quality and processing procedure can be greatly improved for ASKAP.  The addition of more antennas -- to provide more complete sky coverage and achieve higher sensitivity with less integration time -- along with the use of the on-dish calibration system and an overall improved observing method will show vast improvements to the data quality.  With expanded computing systems and a more robust {\sc{ASKAPsoft}} pipeline, larger datasets can be processed more efficiently.  The observations and results presented in this paper serve as one of the first steps for a long and fruitful journey of ASKAP {H\sc{i}} science.

\section*{Acknowledgements}

We thank the reviewer for his/her thoroughly detailed comments and suggestions to improve the clarity of this paper. The Australian SKA Pathfinder is part of the Australia Telescope National Facility which is managed by CSIRO. Operation of ASKAP is funded by the Australian Government with support from the National Collaborative Research Infrastructure Strategy. This work was supported by resources provided by the Pawsey Supercomputing Centre with funding from the Australian Government and the Government of Western Australia, including computational resources provided by the Australian Government under the National Computational Merit Allocation Scheme (project JA3). Establishment of ASKAP, the Murchison Radio-astronomy Observatory and the Pawsey Supercomputing Centre are initiatives of the Australian Government, with support from the Government of Western Australia and the Science and Industry Endowment Fund. We acknowledge the Wajarri Yamatji as the traditional owners of the Observatory site. Parts of this research were conducted by the Australian Research Council Centre of Excellence for All-sky Astrophysics (CAASTRO), through project number CE110001020 as well as by the Australian Research Council Centre of Excellence for All Sky Astrophysics in 3 Dimensions (ASTRO 3D) through project number CE170100013.  P.S. acknowledges funding from the European Research Council (ERC) under the European Union's Horizon 2020 research and innovation programme (grant agreement no. 679627; project name FORNAX).








\appendix

\section*{Appendix: {H\sc{i}} in surrounding group members}

Within the 12-beam image cube, aside from the sources directly associated with the NGC 7232/3 triplet and surrounding tidal debris, there are nine resolved {H\sc{i}} galaxies.  All of these galaxies appear to have stellar counterparts and for those that lie above the HIPASS detection levels, ASKAP recovers similar amounts of flux as HIPASS (see the right column of Figure~A\ref{fig:momA}).  The new {H\sc{i}} detections appear to have fairly smooth velocity gradients that follow the morphology of the assumed stellar counterpart (Figure~A\ref{fig:momB}). The individual characteristics of these nine galaxies are further discussed in this appendix.

\subsubsection*{IC 5171}

\begin{figure*}
\begin{center}
  \includegraphics[width=165mm]{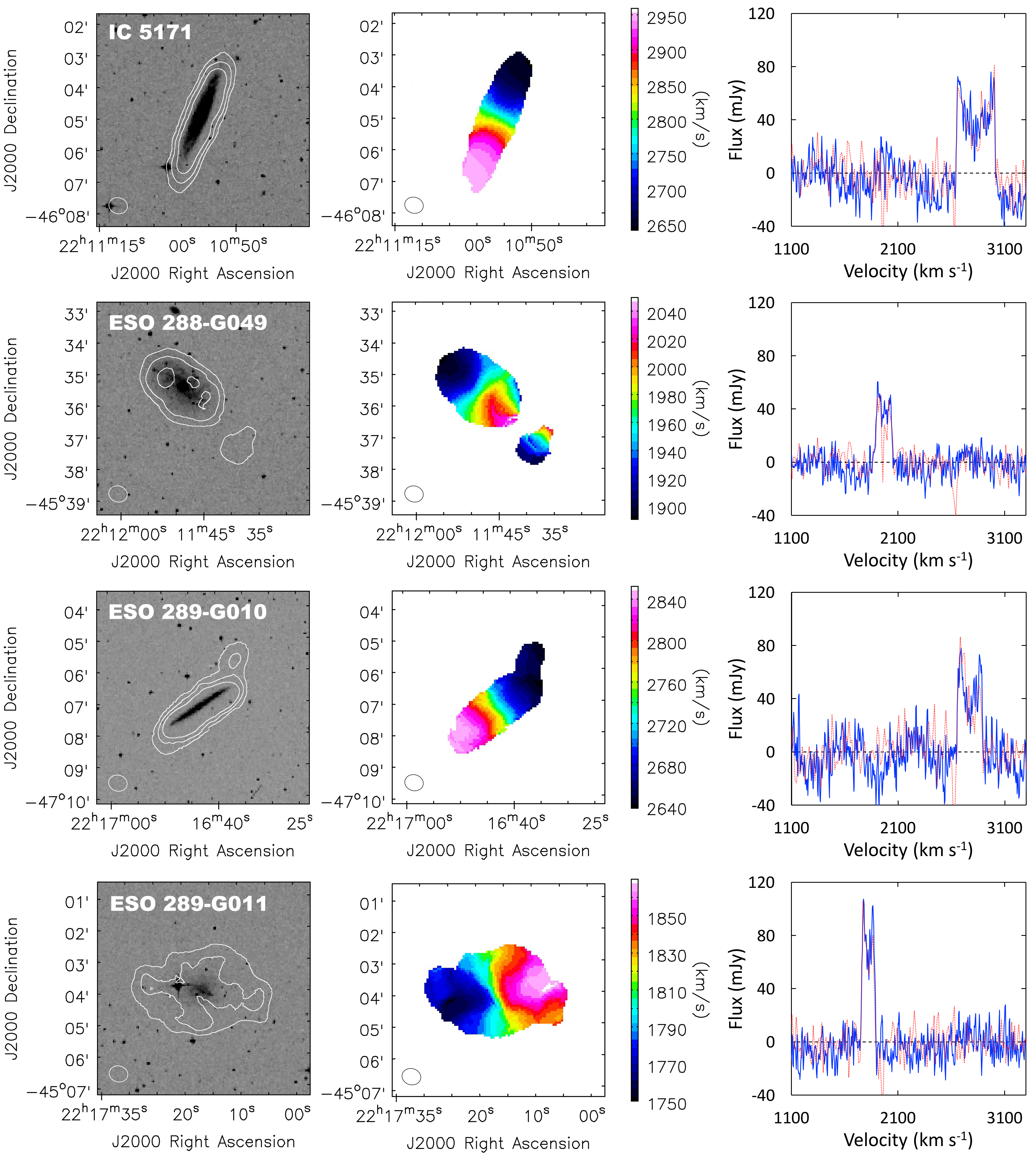}
  \caption{ASKAP {H\sc{i}} detections of galaxies that can be cross-matched to individual HIPASS sources.  Stellar counterparts are labelled and each row represents one source.  For each galaxy we show, left column: the ASKAP {H\sc{i}} integrated intensity (moment 0) contours in white -- at (1, 3, 6) $\times$ 10$^{20}$ atoms cm$^{-2}$ -- superimposed on DSS2 Blue archival images; middle column: ASKAP {H\sc{i}} velocity (moment 1) maps; right column: the ASKAP (solid blue) and HIPASS {H\sc{i}} spectra.  Please note, the residual sidelobe artefacts around IC~5171 have been manually masked out of the moment maps.
  \vspace{20mm}
  \label{fig:momA}}
\end{center}
\end{figure*}

\begin{figure*}
\begin{center}
  \includegraphics[width=160mm]{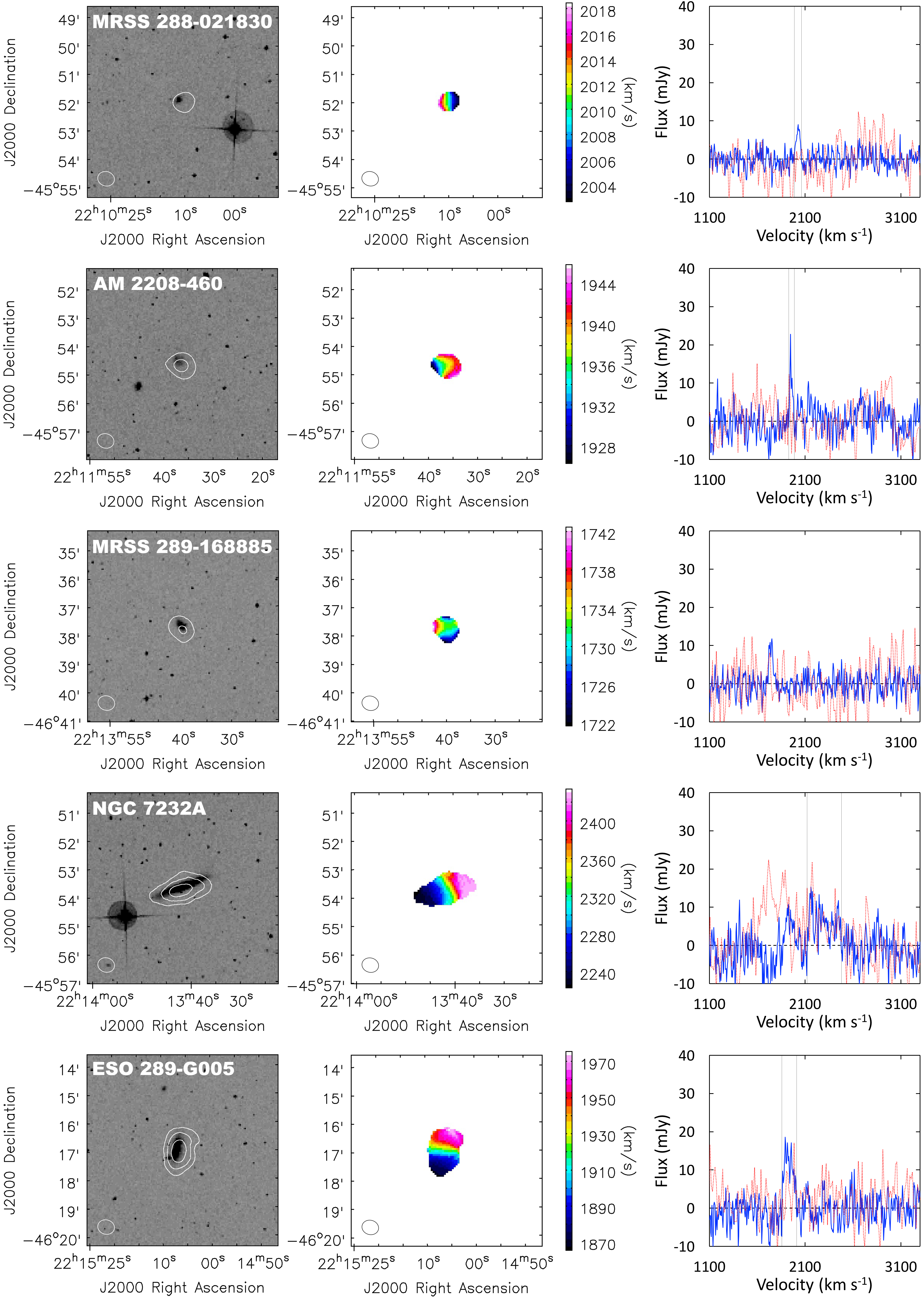}
  \caption{ASKAP {H\sc{i}} detections of galaxies that were hidden within the noise of HIPASS.   Likely stellar counterparts are labelled and each row represents one source.  For each galaxy we show, left column: the ASKAP {H\sc{i}} moment 0 contours in white -- at (1, 3, 6) $\times$ 10$^{20}$ atoms cm$^{-2}$ -- superimposed on DSS2 Blue archival images; middle column: ASKAP {H\sc{i}} moment 1 maps; right column: the ASKAP (solid blue) and HIPASS (dotted red) {H\sc{i}} spectra. The HIPASS spectra were extracted using a single pixel.  Vertical lines indicate the ASKAP {H\sc{i}} velocity ranges of the galaxies.  
\label{fig:momB}}
\end{center}
\end{figure*}

IC~5171 is a bright spiral galaxy that has a high amount of associated {H\sc{i}} emission.  The residual sidelobe artefacts around IC~5171, shown in Figure~\ref{fig:mom0}, are particularly prominent as this galaxy resides in the higher sensitivity region of the {H\sc{i}} cube.  To highlight the actual {H\sc{i}} features of the galaxy, these artefacts have been manually masked out of the moment maps in Figure~A\ref{fig:momA}.  IC~5171 has a higher central velocity than other group members indicating that it is likely in the outskirts of the galaxy group.  Overall, IC~5171 looks fairly symmetrical and undisturbed by the group environment.  

\subsubsection*{ESO 288-G049}

The southwest portion of ESO~288-G049 appears to have a disturbed velocity gradient.  There is also an apparent gas cloud that was picked up by SoFiA as being part of this galaxy.  If the cloud is real (i.e. not a sidelobe / imaging artefact resulting from the preliminary nature of the observations and data processing methods used during the Early Science phases of ASKAP), it could be another {H\sc{i}} source that is interacting with ESO~288-G049.  However, the stellar disk of this face-on spiral and the spectral profile both show no noticeable asymmetries.

\subsubsection*{ESO 289-G010}

ESO 289-G010 is an edge-on spiral located at the edge of the field of view of the 12-beam footprint.  ASKAP recovers the same amount of {H\sc{i}} flux for this galaxy as the HIPASS observations.  Nevertheless, the ripple pattern --  the same artifact that was previously noticed in the beam edge region for some nights of the footprint B observations -- is quite evident in the spectrum for this source.  Although there appears to be an {H\sc{i}} extension to the north of the disk of ESO 289-G010, the RMS noise in this region of the cube is higher than the central regions.  As such, it is unclear if this feature is real.  We only processed and imaged a subset of beams for this paper, the addition of other beams from the same dataset would result in better {H\sc{i}} maps for further investigation of this galaxy.

\subsubsection*{ESO 289-G011}

ESO 289-G011 is optically classified as an irregular galaxy.  Its {H\sc{i}} velocity pattern indicates well-behaved rotation and a velocity width that is common for a typical gas-rich dwarf.  The asymmetries in the outer regions of the galaxy suggest that there is a possible warp in the {H\sc{i}} disk.

\subsubsection*{MRSS 288-021830, AM 2208-460 and MRSS 289-168885}

Our ASKAP observations are the first to detect and resolve the {H\sc{i}} components associated with these sources.  With the high angular and spectral resolution of ASKAP, it is possible to associate these {H\sc{i}} detections with likely stellar counterparts and place constraints on redshift distances. All three galaxies appear to be members of the NGC~7232 galaxy group; however, since their {H\sc{i}} is spatially unresolved, it is difficult to ascertain their dynamical properties or any details about how they are being affected by the group environment.

\subsubsection*{NGC 7232A}

NGC 7232A is a nearly edge-on disk galaxy.  In addition to the the spatial and spectral coincidence between the stellar and {H\sc{i}} emission, the {H\sc{i}} velocity gradient follows the major axis of the stellar disk of NGC~7232A across nearly 200 km s$^{-1}$, confirming that these components belong to the same galaxy.  The spectra for this galaxy appear to be quite noisy, for both ASKAP and HIPASS, but three peaks of emission can be discerned in the ASKAP spectra and image cube.  There is a noticeable feature between $\sim$1600 - 2000 km s$^{-1}$ in the HIPASS spectrum that is not present in the ASKAP spectrum.  This $\sim$2-sigma HIPASS peak modestly coincides with spatial region (within the 15.5 arcmin HIPASS beam) and spectral range of {H\sc{i}} clouds C1 and C2.  If this feature is diffuse {H\sc{i}} emission, it could be part of the tidal bridge between the NGC~7232/3 triplet and IC~5181.

\subsubsection*{ESO 289-G005}

Similar to NGC~7232A, the {H\sc{i}} emission detected in the same spatial and spectral region of ESO 289-G005 has a velocity gradient across the extent of its stellar disk.  However, reflecting the compact morphology of this galaxy, the  {H\sc{i}} spans $\sim$100 km s$^{-1}$.  This detection is the most prominent of the five newly resolved {H\sc{i}} galaxies.

\bsp	
\label{lastpage}
\end{document}